\documentclass[reprint,amsmath,amssymb,aps,showpacs]{revtex4-1}

\usepackage{graphicx}
\usepackage{amsmath}
\usepackage{amssymb}
\usepackage{multirow}
\usepackage{bm}
\usepackage{color}
\usepackage{hyperref}

\usepackage{graphicx}
\usepackage{dcolumn}
\usepackage{bm}

\begin{document}

\title{Absorption features caused by oscillations of electrons
       on the surface of a quark star}

\author{R.~X. Xu$^1$, S.~I. Bastrukov$^{1,2}$, F. Weber$^3$,
  J.~W. Yu$^1$ and I.~V. Molodtsova$^2$\\%
$^1$State Key Laboratory of Nuclear Physics and
Technology, Peking University, Beijing 100871, China\\%
$^2$Joint Institute for Nuclear Research, 141980 Dubna, Russia\\%
$^3$Department of Physics, San Diego State University, San Diego,
    California 92182, USA}

\date{\today}

\begin{abstract}
  If quark stars exist, they may be enveloped in thin electron
  layers (electron seas), which uniformly surround the entire
  star. These layers will be affected by the magnetic fields of quark
  stars in such a way that the electron seas would transmit
  hydromagnetic cyclotron waves, as studied in this paper.  Particular
  attention is devoted to vortex hydrodynamical oscillations of the
  electron sea. The frequency spectrum of these oscillations is
  derived in analytic form. If the thermal X-ray spectra of quark
  stars are modulated by vortex hydrodynamical vibrations, the thermal
  spectra of compact stars, foremost central compact objects (CCOs) and
  X-ray dim isolated neutron stars (XDINSs), could be used to verify the
  existence of these vibrational modes observationally. The central
  compact object 1E 1207.4-5209 appears particularly interesting in
  this context, since its absorption features at 0.7 keV and 1.4 keV
  can be comfortably explained in the framework of the hydro-cyclotron
  oscillation model.
\end{abstract}

\maketitle

\section{Introduction}

The spectral features of thermal X-ray emission are essential for us
to understand the real nature of pulsar-like compact stars.
Calculations show that atomic spectral lines form in the atmospheres
of neutron stars. From the detection and identification of atomic
lines in thermal X-ray spectra one can infer neutron star masses
($M$) and radii ($R$), since the redshift and broadening of the
spectral lines depend on $M/R$ and $M/R^2$, respectively.
Atomic features are expected to be detectable with the spectrographs
on board of {\em Chandra} and {\em XMM-Newton}. No atomic features
have yet been discovered with certainty, however. This may have it
origin in the very strong magnetic fields carried by neutron stars.
An alternative explanation  could be that the underlying compact
star is not a neutron star but a bare strange (quark matter)
star~\cite{Xu02,Xu09}. The surface of such an object does not
consist of atomic nuclei/ions, as it is the case for a neutron star,
but of a sea of electrons which envelopes the quark matter.

Strange stars are quark stars made of absolutely stable strange
quark
matter~\cite{witten,bodmer,terazawa,farhi84:a,schaffner97:a,Weber05}.
They consist of essentially equal numbers of up, down and strange
quarks as well as of electrons
\cite{alcock86:a,alcock88:a,madsen98:b}. The latter are needed to
neutralize the electric charges of the quarks, rendering the
interior of strange stars electrically neutral.
Quark matter is bound by the strong interaction, while electrons are
bound to quark matter by the electromagnetic interaction. Since the
latter is long-range, some of the electrons in the surface region of
a quark star reside outside of the quark matter boundary, leading to
a quark matter core which is surrounded by a fairly thin (thousands
of femtometer thick) sea of electrons
\cite{Weber05,alcock86:a,kettner94:b}.
Due to the enormous advances in X-ray astronomy, more and more
so-called dead pulsars are discovered, whose thermal radiation
dominates over a very weak or negligibly small magnetospheric
activity.
The best absorption features (at $\sim 0.7$ keV and $\sim 1.4$ keV)
were detected for the central compact object (CCO) 1E 1207.4-5209 in
the center of supernova remnant PKS 1209-51/52 (see Table
\ref{table:ccos}). Initially, these
\begin{table}[htb]
\begin{center}
  \caption{Dead pulsars (CCOs and XDINSs) with observed spectral
    absorption lines~\cite{Halpern2010,Turolla2008,CCO2004}, with $P$
    the spin period, $B$ the magnetic field ($B_{10}
    =B/10^{10}$~G) derived by magneto-dipole braking, $T$ the
    effective thermal temperature detected at infinity, and
    $E_{\rm a}$ the absorption energy.  We do not list the $B$-fields
    of XDINSs for which the propeller braking could be significant
    because of their long periods.}
\label{table:ccos} \tabcolsep 4.2pt \small
\begin{tabular}{lllll}
\hline\noalign{\smallskip}
 Source & $P$/s&$B_{10}$& $kT$/keV & $E_{\rm a}$/keV\\
  \hline
  RX J0822.0-4300 &  0.112  & $< 98$& 0.4&$-$ \\
  1E 1207.4-5209  &  0.424 & $< 33$ & 0.22&0.7,~1.4 \\
  CXOU J185238.6+004020  &  0.105  & $3.1$& 0.3&$-$ \\
  RX J0720.4-3125 &8.39&&0.085&0.27\\
  RX J0806.4-4123 &11.37&&0.096&0.46\\
  RX J0420.0-5022 &3.45&&0.045&0.33\\
  RX J1308.6+2127 &10.31&&0.086&0.3\\
  RX J1605.3+3249 &$-$&&0.096&0.45\\
  RX J2143.0+0654 &9.43&&0.104&0.7\\
\noalign{\smallskip}\hline
\end{tabular}
\end{center}
\end{table}
features were thought to be associated with the atomic transitions
of ionized helium in a stellar atmosphere where a strong magnetic
field is present~\cite{1207}. Soon thereafter, however, it was noted
that these lines are of electron-cyclotron origin~\cite{xwq03}.
The spectrum of 1E 1207.4-5209 shows two more features that may be
caused by resonant cyclotron absorption, one at $\sim 2.1$ keV and
another, but of lower significance, at $\sim 2.8$
keV~\cite{Bignami}. These features vary in phase with the star's
rotation.
Although the detailed mechanism which causes the absorption features
is still a matter of debate, timing observations predict a rather
weak magnetic field for this CCO, in agreement to what is obtained
under the assumption that the lowest-energy line at 0.7 keV is the
electron-cyclotron fundamental, favoring the electron-cyclotron
interpretation~\cite{GH07,HG11}.
Besides 1E 1207.4-5209, broad absorption lines have also been
discovered in other dead pulsars (listed in Table
 \ref{table:ccos}), especially in so-called X-ray
  dim isolated neutron stars (XDINSs), between about 0.3 and 0.7
keV~\cite{Turolla2008}.

  In this paper, we re-investigate the physics of these absorption
  features. The key assumption that we make here is that these
  features originate from the electron seas on quarks stars rather
  than from neutron stars, whose surface properties are radically
  different from those of strange stars
  \cite{Weber05,alcock86:a,kettner94:b}. Of key importance is the
  magnetic field carried by a quark star, which critically affects the
  global properties (hydrodynamic surface fluctuations) of the
  electron sea at the surface of the star.
  We study this problem in the framework of classical
  electrodynamics in terms of cyclotron resonances of electrons in
  weak magnetic fields, since the magnetic fields of dead pulsars
  are much lower than the critical field, $B_q=4.414\times
  10^{13}$ G, at which the quantization of the cyclotron orbits of
  electrons into Landau levels occurs.

\section{Hydro-cyclotron waves}

\subsection{Governing equations}

  For what follows we restrict ourselves to a discussion of the
  large-scale oscillations of an electron sea subjected to a stellar
  magnetic field. We will be applying the semi-classical approach of
  classical electron theory of metals and making use of standard
  equations of fluid-mechanics.
  The electrons are viewed as a viscous fluid of uniform density
  $\rho=n\,m_e$ (where $n$ is the electron number density) whose
  oscillations are given in terms of the mean electron flow velocity
  $\delta {\bf v}$.  This implies that the fluctuation
  current-carrying flow is described by the density of the convective
  current $\delta {\bf j}=\rho_e\, \delta {\bf v}$, where
  $\rho_e=e\,n$ is the electron charge density.
The equations of motions of a viscous electron fluid are then given
by~\cite{byp02}
 \begin{eqnarray}
 \label{1.1}
 && \rho\frac{\partial \delta {\bf v}}{\partial t}=\frac{1}{c}
 [\delta{\bf j}\times {\bf B}]+\eta\nabla^2\delta {\bf v},\\ [0.2cm]
 \label{1.2}
 &&
 {\bf j}=\rho_e\delta {\bf v},\quad  \rho=m_e\,n,\quad\rho_e=e\,n \, ,
 \end{eqnarray}
 where $e$ and $n$ are the change and number densities of electrons,
 respectively.
 We emphasize that $\delta {\bf j}$ stands for the convective
 current density and not for Amp\'ere's ${\bf j}=(c/4\pi)\nabla\times
 \delta {\bf B}$, as it is the case for magneto-hydrodynamics. This
 means that the hydrodynamic oscillations in question are of
 non-Alfv\'en type. In Eq.\ (\ref{1.1}), $\eta$ denotes the effective
 viscosity of the electron fluid, which originates from collisions of
 electrons with the magnetic field lines at the stellar surface.
It is worth noting that the cyclotron waves can be regarded as an
analogue of the inertial waves in a rotating incompressible fluid, as
presented in Eq.~(III.56) of Chandrasekhar's
book~\cite{Chandra1961}.

The governing equation, Eq.\ (\ref{1.1}), can be represented as
 \begin{eqnarray}
 \label{e2.1}
 &&\frac{\partial \delta {\bf v}}{\partial t}+
 \omega_c[{\bf n}_B\times \delta {\bf v}]-\nu \nabla^2\delta {\bf
 v}=0,\\ [0.2cm]
 \label{e2.2}
 &&
 \omega_{c}=\frac{eB}{m_e\,c},\quad
{\bf n}_B=\frac{{\bf B}}{B},\quad \nu=\frac{\eta}{\rho}.
 \end{eqnarray}
 where $\omega_{c}$ is the cyclotron frequency.
 In the Appendix, we show that the
 electron sea can transmit macroscopic perturbations in the form of
 rotational hydro-cyclotron waves which are characterized by the
 following dispersion relation,
   \begin{eqnarray}
  \label{e2.3}
  \omega =\pm\,\omega_c\cos\theta\left[\frac{1}{1-(\nu
      k^2/\omega)^2}+i\frac{(\nu k^2/\omega)}{1-(\nu
      k^2/\omega)^2}\right],
  \end{eqnarray}
  where $\omega$ and $k$ denote the frequency and wave vector of the
  perturbations, respectively.
The Larmor radius of an electron in a strong magnetic field, $r_{\rm
L} \simeq m_ec^2/(eB)\propto B^{-1}$, is very small for pulsar-like
compact stars, and we neglect the viscosity term in the following
analysis of the motion of collective electrons.
  In the collision-free regime, $\nu=0$,
  the hydro-cyclotron electron wave is described as a transverse,
  circularly polarized wave whose dispersion relation and propagation
  speed are given by
 \begin{eqnarray}
  \label{e2.4}
 \omega =\pm\omega_c\cos\theta\quad {\rm and}\quad
 V=\pm(\omega_c/k)\cos \theta,
 \end{eqnarray}
 respectively. Here, $\theta$ is the angle between the magnetic field
 ${\bf B}$ and the wave vector ${\bf k}$. If ${\bf k}\parallel {\bf
   B}$ one has $\omega =\pm\omega_c$. In
 metals these kind of oscillations are observed as electron-cyclotron
 resonances. There are two possible resonance states, one for
 $\omega=\omega_c$ and the other for $\omega=-\omega_c$.  These
 resonances correspond to the two opposite orientations of circularly
 polarized electron cyclotron waves.

 \subsection{Hydro-cyclotron oscillations of electrons on bare strange
   quark stars}

We restrict our analysis to the collision-free regime of vortex
hydro-cyclotron oscillations.  Using spherical coordinates, equation
(\ref{e2.1}) then takes the form
\begin{eqnarray}
  \label{e2.5}
  &&\frac{\partial \delta {\bf v}}{\partial t}+
  \omega_c[{\bf n}_B\times \delta {\bf v}]=0.
  \end{eqnarray}
  Taking the curl of both sides of Eq.\ (\ref{e2.5}), we
  obtain
  \begin{eqnarray}
  \label{e2.5a}
 && \frac{\partial \delta \mbox{\boldmath
 $\omega$}}{\partial t}
 =
 \omega_c({\bf n}_B\cdot\nabla)\,\delta {\bf v},\quad \delta
 \mbox{\boldmath $\omega$}=\nabla\times \delta{\bf v}.
  \end{eqnarray}
  Let the magnetic field ${\bf B}$ be directed along the $z$-axis, so
  that in Cartesian coordinates ${\bf n}_B=(0,0,1)$. We then
  have
\begin{eqnarray}
 \label{e2.9}
 && n_r=\cos\theta,\quad n_\theta=-\sin\theta,
\quad n_\phi=0.
\end{eqnarray}
From a mathematical point of view, the problem can be considerably
simplified if one expresses the velocity $\delta{\bf v}$, which
obeys the condition $\nabla\cdot \delta{\bf v}=0$, in terms of
Stokes' stream function, $\chi(\theta,\phi)$. This leads to
\begin{eqnarray}
 \label{e2.10}
 \delta v_r=0,\, \delta v_\theta=\frac{1}{r\sin\theta}\frac{\partial
   \chi(\theta,\phi)}{\partial
   \phi},\, \delta v_\phi
 =-\frac{1}{r}\frac{\partial \chi(\theta,\phi)}{\partial \theta}.
\end{eqnarray}
The depth of the electron layer near the star is much smaller than
the stellar radius so that $r\approx R$ to a very good
approximation. Equation (\ref{e2.5a}) then simplifies to
\begin{eqnarray}
 \label{e2.11}
 &&\frac{\partial \delta \omega_r}{\partial t}
 =-\omega_c\,\frac{n_\theta \delta
   v_\theta}{R},
  \end{eqnarray}
  with the radial component of the vortex given in terms of
  $\chi$,
\begin{eqnarray}
 \label{e2.12}
 \delta \omega_r=
 \frac{1}{R}\left[\frac{\partial \delta v_\phi}{\partial \theta}-
 \frac{1}{\sin\theta}\frac{\partial \delta v_\theta}{\partial \phi}\right]=
 -\frac{1}{R^2}
  \nabla^2_{\perp}\chi(\theta,\phi),\\
  \label{e2.13}
\nabla_{\perp}^2=
  \frac{1}{\sin\theta}\frac{\partial }{\partial \theta}
  \left(\sin\theta\frac{\partial }{\partial \theta}\right)+\frac{1}{\sin^2\theta}
  \frac{\partial^2 }{\partial \phi^2}.\quad\quad\,
  \end{eqnarray}
  Substituting Eq.\ (\ref{e2.12}) and Eq.\ (\ref{e2.13}) into
  Eq.\ (\ref{e2.11}) leads to
 \begin{eqnarray}
 \label{e2.14}
  \nabla_{\perp}^2\left
 (\frac{\partial \chi}{\partial t}\right)
 +\omega_c\frac{\partial \chi}{\partial \phi}=0.
  \end{eqnarray}
  The fact that free electrons undergo cyclotron oscillations in the
  planes perpendicular to the magnetic field suggests that the stream
  function $\chi$ can be written in the following separable form,
 \begin{eqnarray}
 \label{e2.15}
  \chi(\theta,\phi)=\psi(\theta)\cos(\phi\pm\omega t).
  \end{eqnarray}
  The ``$+$'' sign allows for cyclotron oscillations which are induced
  by the clockwise polarized wave, and the ``$-$''
  sign allows for cyclotron oscillations induced by the
  count-clockwise polarized wave. Substituting (\ref{e2.15})
  into (\ref{e2.14}) leads to
 \begin{eqnarray}
 \label{e2.16}
  \nabla_{\perp}^2\,\psi(\theta)\pm\frac{\omega_c}{\omega}
 \psi(\theta)=0.
  \end{eqnarray}
  In the reference frame where the polar axis is fixed,
  Eq.\ (\ref{e2.16}) is identical to the Legendre equation for
  the surface spherical function,
\begin{eqnarray}
 \label{e2.17}
  \nabla_{\perp}^2\,P_{\ell}(\theta)+\ell(\ell+1)P_{\ell}(\theta)=0,
  \end{eqnarray}
  where $P_{\ell}(\cos\theta)$ denotes the Legendre polynomial of
  degree $\ell$. Hence, setting $\psi(\theta)=P_{\ell}(\theta)$ we
  obtain
\begin{eqnarray}
 \label{e2.18}
 \omega_\pm(\ell)=\pm
 \frac{\omega_c}{\ell(\ell+1)},\quad
 \omega_{c}=\frac{eB}{m_ec},\quad \ell\geq 1.
\end{eqnarray}
From this relation we can read off the frequency of a surface
hydro-cyclotron oscillation of a given order $\ell$.

\subsection{Characteristic features of cyclotron frequencies}

Let us consider the spectrum of the positive branch
$\omega(\ell)=\omega_{+}(\ell)$ of Eq.\ (\ref{e2.18}),
 \begin{eqnarray}
 \label{a2.14}
 &&\frac{\omega(\ell)}{\omega_c}=\frac{1}{\ell(\ell+1)},\quad \ell\geq 1.
  \end{eqnarray}
From
\begin{eqnarray}
 \label{a2.15}
 &&\frac{\omega(\ell)}{\omega(\ell+1)}=\frac{\ell+2}{\ell},\quad \ell\geq
 1,
\end{eqnarray}
it follows that this ratio becomes a constant for $\ell\gg 1$.
Such a spectral feature is notably different from the one of
electron-cyclotron resonances of transitions between different
Landau levels.

From the energy eigenvalues, $E_n$, of an electron in a strong
magnetic field, which are found by solving the Dirac equation (see
Ref.\ \cite{Bussard}), one may approximate the value of $E_n$ for a
relatively weak magnetic field, $B\ll B_q$, by
\begin{eqnarray}
 E_n=mc^2+n \hbar\omega_c,\,\, n\geq 0.
\end{eqnarray}
Therefore, in the framework of a single-particle approximation, the
emission/absorption frequencies, which are given by
  $\hbar\omega(\ell) = E_{n+\ell}-E_n$, should occur at
\begin{eqnarray}
 \label{landau}
 &&\omega(\ell)=\ell\omega_c,\quad \ell\geq 1.
\end{eqnarray}
Table \ref{tab:1} compares the results of Eq.\ (\ref{landau}) with
the results of Eq.\ (\ref{a2.14}) obtained for the hydro-cyclotron
wave model.
Most notably, it follows that for the hydro-cyclotron wave model one
obtains $\omega(\ell=2)/\omega(\ell=3)=2$, in contrast to the
cyclotron resonance model for single electrons which predicts this
ratio for $\omega(\ell=2)/\omega(\ell=1)=2$..
\begin{table}
  \caption{Comparison between single-particle (Landau level) and
    hydro-wave results for the cyclotron frequencies ($B_{12}
    =B/10^{12}$~G, $\omega_1=\omega(\ell=1)$, $\omega_c$ denotes
    the cyclotron frequency).}
\label{tab:1}
\begin{tabular}{lllll}
  \hline\noalign{\smallskip}
  ~ & $\omega(\ell=1)$ & $\omega(\ell=2)$ & $\omega(\ell=3)$ &
  $\hbar\omega_1$/keV \\
  \noalign{\smallskip}\hline\noalign{\smallskip}
  Landau level & $\omega_c$ & $2\omega_c$ & $3\omega_c$ & $11.6B_{12}$ \\
  Hydro-wave & $\omega_c/2$ & $\omega_c/6$ & $\omega_c/12$ & $5.8B_{12}$\\
  \noalign{\smallskip}\hline
\end{tabular}
\end{table}

\section{1E 1207.4-5209 and other compact objects}

As already mentioned in the Introduction, 1E 1207.4-5209 (or
J1210-5226) in PKS 1209-51/52 is one of the central compact objects
in supernova remnants~\cite{CCO2004}, where broad absorption lines,
near (0.7, 1.4) keV~\cite{1207}, and possibly near (2.1, 2.8)
keV~\cite{Bignami} were detected for the first time.
The interpretation of the absorption feature at $\sim 2.8$ keV is
currently a matter of debate, in contrast to the feature at $\sim
2.1$ keV which is essentially unexplained. Intriguingly, an
  absorption feature with the same energy, 2.1 keV, has also been
detected in the accretion-driven X-ray pulsar 4U
1538-52~\cite{2.1keV}.

  For what follows, we assume that 1E 1207.4-5209 is a strange
  quark star and that (some of) these absorption features are produced
  by the hydro-cyclotron oscillations of the electron sea at the
  surface of such an object.  Assuming a magnetic surface field of
  $B\simeq7\times 10^{11}$ G and thus $\omega(\ell=3)=0.7$ keV, we obtain
  the oscillation frequencies shown in Table \ref{tab:1207}.
\begin{table}
  \caption{The frequencies, $\omega(\ell)$, at which hydro-cyclotron
    oscillations occur for 1E 1207.4-5209 with effective temperature $T\simeq 0.2$
    keV, assuming a  magnetic field of $B\simeq 7\times 10^{11}$ G.}
\label{tab:1207}
\begin{tabular}{lllllll}
\hline\noalign{\smallskip}
$\ell$ & $1$ & $2$ & $3$ & $4$ & $5$ & $6$\\
\noalign{\smallskip}\hline\noalign{\smallskip}
$\omega(\ell)$/keV~~~ & 4.2~~~ & 1.4~~~ & 0.7~~~ & 0.4~~~ & 0.3~~~ & 0.2\\
\noalign{\smallskip}\hline
\end{tabular}
\end{table}
  A magnetic field of $\sim 7\times 10^{11}$ G is compatible with
  the magnetic fields inferred for 1E 1207.4-5209 from timing
  solutions~\cite{HG11} ($9.9\times 10^{10}$ G or $2.4\times 10^{11}$
  G), since 1E 1207.4-5209 shows no magnetospheric activity and the
  $\dot P$-value would be overestimated if one applies the spin-down
  power of magnetic-dipole radiation~\cite{xq01,k06}.  We note that
  the absorption feature at $\omega(\ell=1)= 4.2$ keV shown in Table
  \ref{tab:1207} may not be detectable since the stellar temperature
  is only $\sim 0.2$ keV (see Table \ref{table:ccos}), which will
  suppress any thermal feature in that energy range.

Aside from 1E 1207.4-5209, one may ask what would be the
  magnetic fields of other dead pulsars (e.g.,
  radio-quiet compact objects) if their
  spectral absorption features would also be of hydro-cyclotron
  origin? Intriguingly, the hydro-cyclotron wave model predicts
  magnetic fields that are twice as large as those derived from the
  electron cyclotron model if the absorption feature is at
  $\omega(\ell=1)=\omega_c/2$; these fields could be $\sim 10$ times
  greater (see Table \ref{tab:1}) if the absorption feature is at
  $\omega(\ell=2)=\omega_c/6$ or $\omega(\ell=3)=\omega_c/12$.
The absorption lines at $(0.3\sim 0.7)$ keV may indicate that the
fields of XDINSs are on the order of $\sim 10^{10}$ to $10^{11}$ G,
if oscillation modes with $\ell\geq 4$ are not significant.

As noted in \cite{xwq03}, unique absorption features on compact
stars are only detectable with {\em Chandra} and {\em XMM-Newton} if
the stellar magnetic fields are relatively weak ($10^{10}$ G to
$10^{11}$ G), since the stellar temperatures are only a few 0.1 keV.
The fields of many pulsar-like objects are generally greater than
this value, with the exception of old millisecond pulsars whose
fields are on the order of $10^8$ G. Central compact objects, on the
other hand, seem to have sufficiently weak magnetic fields (see
Table \ref{table:ccos}) so that absorption features originating from
their surfaces should be detectable by {\em Chandra} and {\em
  XMM-Newton}. Arguments favoring the interpretation of compact
central objects as strange quark stars have been put forward in
\cite{negreiros:10cco}, where it was shown that the magnetic field
observed for some CCOs could be generated by small amounts of
differential rotation between the quark matter core and the electron
sea.

Besides dead pulsars, anomalous X-ray pulsars (AXPs) and soft
gamma-ray repeaters (SGRs) are enigmatic objects which have become
hot topics of modern astrophysics.
Whether they are magnetars/quark stars is an open
question~\cite{tx11}.
In case that AXPs/SGRSs should be bare strange stars, the absorption
lines detected from SGR 1806-20 could be understood in the framework
of the hydro-cyclotron oscillation model.
\begin{table}
  \caption{Comparison between the frequencies, $\omega(\ell)$, and the
  absorption frequencies detected, $\omega_{\rm obs}$, for SGR 1806-20
  (with an assumption of $B\simeq 1.86\times 10^{13}$ G). The data of
  $\omega_{\rm obs}$ are from \cite{Ibrahim02}. Both $\omega(\ell)$ and
  $\omega_{\rm obs}$ are in keV.}
\label{tab:1806}
\begin{tabular}{lllllllll}
\hline\noalign{\smallskip}
$\ell$ & $1$ & $2$ & $3$ & $4$ & $5$ & $6$ & $7$ & $8$\\
\noalign{\smallskip}\hline\noalign{\smallskip}
$\omega(\ell)$~ &108~ & 36 & 18 & 10.8 & 7.2 & 5.1 & 3.8 & 3.0\\
$\omega_{\rm obs}$~ & $-$ & $-$ & $17.5\pm .5$ & $11.2\pm .4$ & $7.5\pm .3$ & $5.0\pm .2$ & $-$ & $-$\\
\noalign{\smallskip}\hline
\end{tabular}
\end{table}
Assuming a normal magnetic field of $B\simeq 1.86\times 10^{13}$ G
so that $\hbar\omega_c/12=18$ keV, one sees that the oscillation
model predicts hydro-cyclotron frequencies which coincide with the
observed listed in Table~\ref{tab:1806}.

\section{Summary}

In this paper, we study the global motion of the electron seas on
the surfaces of hypothetical strange quark stars.  It is found that
such electron seas may undergo hydro-cyclotron oscillations whose
frequencies are given by $\omega(\ell)=\omega_c/[\ell(\ell+1)]$,
where $\ell \geq 1$ and $\omega_c$ the cyclotron frequency.
We propose that some of the absorption features detected in the
thermal X-ray spectra of dead (e.g., radio silent) compact objects
may have their origin in excitations of these hydro-cyclotron
oscillations of the electron sea, provided these stellar objects are
interpreted as strange quark stars.
The central compact object 1E 1207.4-5209 appears particularly
interesting. It shows an absorption feature at 0.7 keV which is not
much stronger than the another absorption feature observed at 1.4
keV. This can be readily explained in the framework of the
hydro-cyclotron oscillation model, since two lines with $\ell$ and
$\ell+1$ could essentially have the same intensity. This is very
different for the electron-cyclotron model, for which the oscillator
strength of the first harmonic is much smaller than the oscillator
strength of the fundamental.

\section*{Appendix}

Here we derive the dispersion relation characterizing the
propagation of hydro-cyclotron electron wave in the slab-geometry
approximation.
The governing equation of viscous electron fluid under the action of
Lorentz force is given by
$$
\rho\frac{\partial \delta {\bf v}}{\partial t}=\frac{\rho_e}{c}
 [\delta{\bf v}\times {\bf B}]+\eta\nabla^2\delta {\bf v},
$$
which can be written as
\begin{eqnarray}
 \label{a1.1}
 && \frac{\partial \delta {\bf v}}{\partial t}+
 \omega_c[{\bf n}_B\times \delta {\bf v}]-\nu \nabla^2\delta {\bf
   v}=0,
\end{eqnarray}
where
$$
\omega_{c}=\frac{eB}{m_e\,c}, \quad{\bf n}_B=\frac{{\bf
    B}}{B},\quad\nu=\frac{\eta}{\rho},\quad \rho=m_en,\quad\rho_e=en,
$$
$\omega_{c}$ is the cyclotron frequency, and $\eta$ stands for the
effective viscosity of an electron fluid originating from collisions
between electrons. To make the problem analytically tractable, we
treat the electron sea as an incompressible fluid and assume a
uniform magnetic field. Equation (\ref{a1.1}) can then be written as
\begin{eqnarray}
 \label{a1.3}
\frac{\partial \delta {\bf v}}{\partial t}=-\omega_{c}
 [{\bf n}_B\times \delta {\bf v}]+\nu\,\nabla^2\delta {\bf v}.
\end{eqnarray}
%
%
  Upon applying to Eq.\ (\ref{a1.3}) the operator $\nabla\times$, we
  arrive at
\begin{eqnarray}
 \label{a1.4}
 \frac{\partial }{\partial t}[\nabla\times \delta{\bf v}]
 =
 \omega_c({\bf n}_B\cdot\nabla)\,\delta {\bf v} -
 \nu\,\nabla\times\nabla\times\nabla\times \delta {\bf
   v},
\end{eqnarray}
where $\nabla\cdot \delta {\bf v}=0$, and
\begin{eqnarray}
  \label{a1.5}
  && \frac{\partial \delta \mbox{\boldmath
      $\omega$}}{\partial t}
  =
  \omega_c({\bf n}_B\cdot\nabla)\,\delta {\bf v} - \nu \nabla\times\nabla\times
  \delta \mbox{\boldmath $\omega$},
\end{eqnarray}
where $\delta \mbox{\boldmath $\omega$}=\nabla\times \delta{\bf v}$.
Considering a perturbation in the form of $\delta {\bf v}={\bf
v}'\exp[i({\bf kr}-\omega t)]$, we have
\begin{eqnarray}
  \label{a1.6}
  && [{\bf k}\times \delta {\bf v}]=i\frac{\omega_c}{\omega} ({\bf
    n}_B\cdot {\bf k}) \delta {\bf v}+i\frac{\nu k^2}{\omega} [{\bf k}\times \delta
  {\bf v}],\\[0.2cm]
   \label{a1.7}
   && \left(1-i\frac{\nu k^2}{\omega}\right)[{\bf k}\times \delta {\bf
     v}]=\,i\frac{\omega_c}{\omega} ({\bf
     n}_B\cdot {\bf k}) \delta {\bf v},
\end{eqnarray}
where $({\bf k}\cdot \delta {\bf v})=0$. It is convenient to rewrite
the last equation as
\begin{eqnarray}
 \label{a1.8}
 [{\bf k}\times \delta {\bf
   v}]=\,i\frac{\omega_c}{\omega} ({\bf
   n}_B\cdot {\bf k})\,\delta {\bf v}
 \left[\frac{1+i(\nu k^2/\omega)}{1-(\nu k^2/\omega)^2}\right].
\end{eqnarray}
Multiplication of both sides of Eq.\ (\ref{a1.8}) with ${\bf k}$
leads to
\begin{eqnarray}
  \label{a1.9}
  -\omega \delta {\bf v} k^2=i \omega_c ({\bf n}_B\cdot {\bf
    k})\left[\frac{1+i(\nu k^2/\omega)}{1-(\nu k^2/\omega)^2}\right]
  [{\bf k}\times \delta {\bf v}].
\end{eqnarray}
Inserting the left-hand-side of Eq.\ (\ref{a1.8}) into the
right-hand-side of Eq.\ (\ref{a1.9}) gives
\begin{eqnarray}
  \label{a1.10}
  && \omega^2 =\omega_c^2 \frac{({\bf n}_B\cdot {\bf
      k})^2}{k^2}\left[\frac{1+i(\nu k^2/\omega)}{1-(\nu
      k^2/\omega)^2}\right]^2,
\end{eqnarray}
or
$$
\omega =\pm\omega_c \frac{({\bf n}_B\cdot {\bf
    k})}{k}\left[\frac{1+i(\nu k^2/\omega)}{1-(\nu
    k^2/\omega)^2}\right],
$$
which is Eq.\ (\ref{e2.3}).

\vspace{1mm}

{\bf Acknowledgement.}
We would like to acknowledge valuable discussions at the PKU pulsar
group. F.W. is supported by the National Science Foundation (USA)
under Grant PHY-0854699. This work is supported by the National
Natural Science Foundation of China (grants 10935001 and 10973002),
the National Basic Research Program of China (grants 2009CB824800,
2012CB821800), and the John Templeton Foundation.

\end{document}